\documentstyle[11pt,paspconf,epsf]{article}

\def\lya{Ly$\alpha$\ }

\def\kms{{\rm\,km\,s^{-1}}}

\def\hmpc{h^{-1}{\rm\,Mpc}}
\def\hkpc{h^{-1}{\rm\,kpc}}
\def\msun{{\,M_\odot}}

\def\frac#1#2{{#1 \over #2}}
\newbox\grsign \setbox\grsign=\hbox{$>$} \newdimen\grdimen \grdimen=\ht\grsign
\newbox\simlessbox \newbox\simgreatbox
\setbox\simgreatbox=\hbox{\raise.5ex\hbox{$>$}\llap
     {\lower.5ex\hbox{$\sim$}}}\ht1=\grdimen\dp1=0pt
\setbox\simlessbox=\hbox{\raise.5ex\hbox{$<$}\llap
     {\lower.5ex\hbox{$\sim$}}}\ht2=\grdimen\dp2=0pt
\def\simgt{\mathrel{\copy\simgreatbox}}
\def\simlt{\mathrel{\copy\simlessbox}}
\def\gta{\simgt}
\def\lta{\simlt}

\begin{document}

\title{Simulating Gas at High Redshift}

\author{Neal Katz}
\affil{Department of Physics and Astronomy, University of Massachusetts,
    Amherst, MA 01003}

\author{Lars Hernquist}
\affil{Lick Observatory, University of California, Santa Cruz, CA 95064}

\author{David H. Weinberg}
\affil{Department of Astronomy, The Ohio State University, Columbus, OH 43210}

\begin{abstract}
We discuss simulations of gas at high redshift.  We briefly review
the methods used and the results for quasar absorption lines.  We present
gas mass functions and galaxy correlation functions for 5 different
cosmological models.  Galaxies should be detectable at redshifts greater
than 2 by SKAI, and measurements of the gas mass functions and galaxy
correlation functions could be used to discriminate between different
cosmological models.
\end{abstract}

\section{Introduction}

Given all the talks in this conference about proposed telescopes to
detect gas at high redshifts it is comforting as a theorist to think that
here is an area where the theory might be ahead of the observations for a
change.  With the development of sophisticated computer codes and
high performance computers it is now possible to model {\it ab initio } theories
of structure formation with enough accuracy to make detailed predictions.
In particular, unlike the computer simulations of the last decade, it is now
possible to model the gas directly, including many important physical effects
such as radiative cooling and heating by a photoionizing background field.

The latter effect is particularly important when modeling the \lya forest.
After all, observations of quasar absorption lines are observations of
gas at high redshift, albeit at optical not radio or mm wavelengths.  As we
will discuss in this paper, it is our success in matching the observations
of quasar absorption lines that gives us the confidence to extend our work
to predicting radio and mm emission from high redshift gas.
Before proceeding to the following sections, where we describe the methods used
in the simulations, the simulations themselves, comparisons with observed 
quasar absorption lines, and possibilities for observing the gas in emission,
we would like to quickly review some of the basic questions we hope to answer
by comparing theories with observations of gas at high redshifts.

What were the physical conditions of the primordial universe?
What fraction of the matter was in a diffuse medium and what fraction
in dense clouds?  When did dense, neutral clouds first form? 
What fraction and types of dark
matter were there? When and how did the formation of galaxies and large scale
structure begin? How many metals were produced and how early?
What was the typical background radiation field, how homogeneous was it, and
what was producing it? Does the standard big bang model make the correct
predictions about primordial element abundances?
\section{Methods and Models}

We perform our simulations using TreeSPH
(Hernquist \& Katz 1989), a code that unites smoothed particle
hydrodynamics (SPH; Lucy 1977; Gingold \& Monaghan 1977) with the
hierarchical tree method (Barnes \& Hut 1986) for computing
gravitational forces.  The collisionless matter and the gas are both
represented by particles; collisionless particles are influenced only
by gravity, while the gas is subject to gravitational forces, pressure
gradients, and shocks. We include the effect of both radiative cooling,
assuming primordial abundances, and Compton cooling.  We also include
the ionization and heat input from a UV radiation background under the
optically thin assumption, using the field computed by Haardt \& Madau (1996).
Star formation is included heuristically to turn cold, dense gas into
collisionless particles. Stars are allowed to form only in regions that are
in convergent parts of the flow and that are locally Jeans unstable.
Supernova heating is added to the remaining gas assuming a standard IMF from
$0.1$ to $100 {\rm M}_\odot$ and that stars above $8 {\rm M}_\odot$ become
supernovae.  Each supernova adds $10^{51}$ ergs of heat to the system.
The method is described in detail in Katz, Weinberg, \& Hernquist (1996).

Because it uses a Lagrangian hydrodynamics algorithm and
individual particle time steps, TreeSPH can perform simulations
with the enormous dynamic range essential for cosmological applications.
To study high redshift galaxies such a large dynamic range is particularly 
important.  In SPH, gas properties are computed by
averaging or ``smoothing'' over a fixed number of neighboring
particles, typically 30--100.  When matter is distributed
homogeneously, all particles have similar smoothing volumes.  However,
smoothing lengths in TreeSPH are allowed to decrease in collapsing
regions, in proportion to the interparticle separation, thus increasing
the spatial resolution in precisely those regions where a high
dynamic range is needed.  In underdense regions, the smoothing lengths are
larger, but this is physically reasonable because the gas
distribution {\it is} smoother in these regions, requiring fewer
particles for an accurate representation.  TreeSPH allows particles to
have individual time steps according to their physical state, so that
the pace of the overall computation is not driven by the small
fraction of particles requiring the smallest time steps.  The simulations
presented here were performed using a highly vectorized but
serial version of the code. We now have a version that runs on many
parallel machines (Dav\'e, Dubinski \& Hernquist 1997),
allowing us to perform even larger simulations in the future.

Here we present the results of 5 simulations of 5 different models. Each
simulation, of a periodic cube that is $11\hmpc$ on a side (where
$h \equiv H_{0}/100\kms$ Mpc$^{-1}$),
uses $2 \times 64^3$ particles and is evolved to $z=2$.
Each has a nominal gas mass resolution (32 gas particles) of
$4.7 \times 10^{9} (\Omega_b/0.05)\msun$ and a spatial resolution (gravitational
softening length, in physical units) of $6(1 + z)^{-1} \hkpc$.
They all use $\Omega_b  = 0.0125 h^{-2}$ to satisfy the nucleosynthesis 
constraint (Walker et al. 1991).

The first model is ``standard'' CDM (SCDM), with $\Omega=1$, $h=0.5$.
The power spectrum is normalized so that the rms
amplitude of mass fluctuations in $8\hmpc$ spheres, linearly
extrapolated to $z=0$, is $\sigma_{8}=0.7$.
This normalization is consistent with that advocated by White, Efstathiou 
\& Frenk (1993) to match the observed masses of rich galaxy clusters,
but it is inconsistent with the normalization implied by the
COBE-DMR experiment.  Our second model (CCDM) is identical to the first except 
that $\sigma_{8}=1.2$, consistent with the 4-year COBE data 
(Bennett et al.\ 1996).  However, it produces rich clusters that are
too massive.  The third model, OCDM, assumes an open universe with
$\Omega_{0}=0.4$, $h=0.65$, and is also COBE-normalized (Ratra et al.\ 1997)
and produces clusters of about the right mass.
The fourth model, a nonzero-$\Lambda$ CDM model (LCDM) has a tilted initial
power spectrum to match both COBE and cluster masses. The fifth model (TCDM)
is an $\Omega=1$ model that also has a tilted initial power spectrum to 
match both COBE and cluster masses.  The parameters of all the models are
given in Table 1.

\begin{table}
\caption{Models} \label{tbl-1}
\begin{center}\scriptsize
\begin{tabular}{crrrrrrrrrrr}
Model & $\Omega$ & $\Lambda$ & $H_0$ & $\Omega_b$ & $\sigma_{8}$ & $n$ \\
\tableline
SCDM  &     1.0  &    0.0  &     50  &    0.05  &    0.7   &   1.0 \\
CCDM  &     1.0  &    0.0  &     50  &    0.05  &    1.2   &   1.0 \\
OCDM  &     0.4  &    0.0  &     65  &    0.03  &    0.75  &   1.0 \\
LCDM  &     0.4  &    0.6  &     65  &    0.03  &    0.8   &   0.93 \\
TCDM  &     1.0  &    0.0  &     50  &    0.05  &    0.54  &   0.80 \\

\end{tabular}
\end{center}
\end{table}

\section{The Simulations}

At high redshift the gas ends up in three main components:
low density, highly ionized
gas with $\rho/\bar{\rho} \lta 10$ and $T \lta 10^5$ K,
shock heated gas with typical overdensity $\rho/\bar{\rho} \sim 10$--$10^4$
and $T \sim 10^5$--$10^7$ K, and
radiatively cooled, dense gas with $\rho/\bar{\rho} \gta 1000$ and
$T \sim 10^4$ K, as seen in the 
left panel of Figure 1.  This panel shows the distribution of the gas
particles in the temperature-density plane for SCDM at $z = 2$.
The first component gives rise to the \lya forest and follows the
relationship $T = T_0 (\rho /\bar{\rho})^\gamma$ with
$T_0 \approx 6000$ K and $\gamma \approx 0.6$, as shown in the right panel
of Figure 1.  The second component is 
associated with galaxy halos, Lyman limit and metal line systems, and 
intracluster gas, while the third component is associated with damped \lya
systems and high redshift galaxies.

\begin{figure}
\plotone{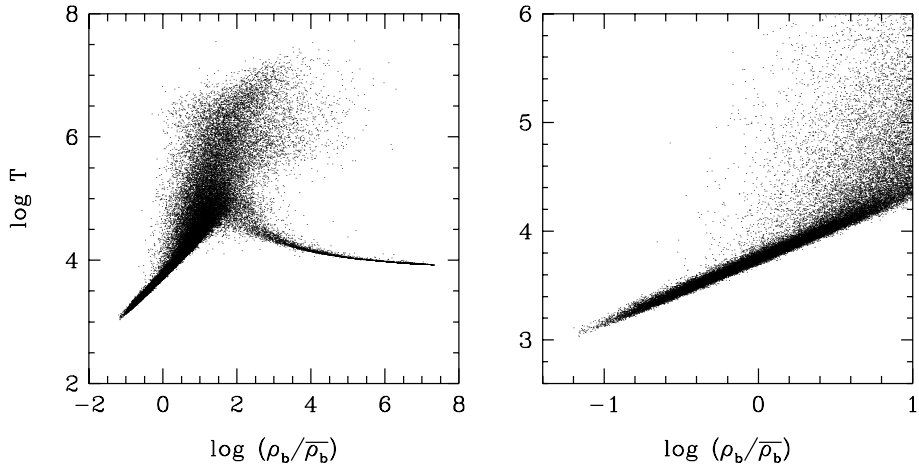}
\caption{The distribution of gas particles in the temperature-density plane.} 
\label{fig-1}
\end{figure}

We can use the simulations to make artificial absorption spectra (Hernquist
et al. 1996) to compare with the observations.  ``Typical'' low column density
absorbers --- to the extent that we can identify such
a class --- are flattened structures of rather low overdensity
($\rho/\bar\rho \sim 1-10$), and have $b$-parameters that are often
set by peculiar motions or Hubble flow rather than thermal broadening.
Many systems are
still expanding with residual Hubble flow, so their physical densities
and neutral fractions decrease with time. The evolution of the \lya
forest is driven primarily by the
increase in physical density with $z$, which raises the neutral
fraction, and hence the opacity, of individual absorbers.  
Traditional searches for the Gunn-Peterson effect implicitly
assume a uniform intergalactic medium (IGM) punctuated by discrete
clouds; thus, absorption in identified lines is removed before seeking
a continuum depression.  The simulations
reveal a smoothly fluctuating IGM, with no sharp distinction between
``background'' and ``\lya clouds''.  One might even say that the 
\lya forest {\it is} the Gunn-Peterson effect.
The \lya forest contains most of the baryons in the Universe at $z=2-5$.

As seen in Figure 2, for the SCDM model, the number of absorption systems
matches the observations for systems with column densities from
$10^{12.5}$ cm$^{-2}$ to $10^{17}$ cm$^{-2}$ (Hernquist et al. 1996; 
Dav\'e et al. 1997).  Dav\'e et al. (1997) also show that 
the velocity widths of the \lya absorbers match the observed
distribution of line widths.  All the models presented here also match these
observations.  The ``standard'' cold dark matter model and
some of its variants can also naturally explain the associated HeII absorption
(Croft et al. 1997) and metal-line absorption (Hellsten et al. 1997).

\begin{figure}
\plotfiddle{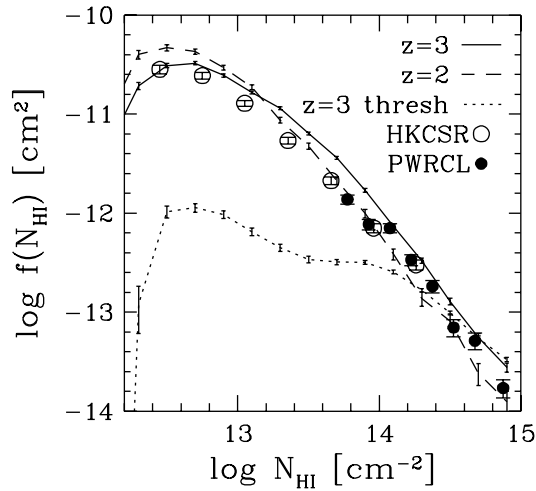}{2.5truein}{0}{75}{75}{-150}{-340}
\caption{The column density distributions $f(N_{HI})$,
the number of lines per unit
redshift per linear interval of HI column density.  Solid and dashed lines show
the simulation results at $z = 3$ and $z = 2$, respectively.  Filled and open
circles show the observational results of Petitjean et al. (PWRCL;1993) and
Hu et al.  (HKCSR;1995), respectively.  From Dav\'e et al.\ (1997).} 
\label{fig-2}
\end{figure}

The high column density systems are associated with galaxies.  Damped \lya
absorbers, $N \geq 10^{20.2}$ cm$^{-2}$, are lines of sight that pass
near normal galaxies.  Lyman limit absorbers ($10^{17}$ cm$^{-2} \leq N \leq 
10^{20.2}$ cm$^{-2}$) are lines of sight that pass through the halos of normal
galaxies or near small dwarf galaxies. After a correction is made for self
shielding, the number of high column density absorption systems 
in the SCDM model matches the 
observations for systems $ >10^{20}$ cm$^{-2}$ (Katz et al. 1996;
Gardner et al. 1997ab).
The CCDM, OCDM, and LCDM models are roughly consistent with the
observed damped \lya abundance, but the TCDM model has too little
small scale power and fails to produce enough damped absorption.

In the future, we plan to make more detailed comparisons with observations.
In particular, we plan to analyze the simulated and observed spectra in the
same exact way and compare the $f(N)$'s, $b$'s, and line coincidences along
closely spaced lines of sight.  We plan to investigate the effects of our
limited resolution, examine more cosmological models, and simulate the
\lya forest at low redshift.

Using the simulations we have shown that {\it ab initio} theories
of structure formation, like CDM, make robust predictions and appear to match
the properties of observed \lya absorption systems.  Hence, we can use
the detailed physical state of the gas to calculate other observable
properties of the gas.

In Figure 3 we plot the cumulative gas mass function of the galaxies in
each of the five models at $z = 2$, 3, and 4.  This mass does not
include the stellar or dark matter components. We
include all galactic gas whose temperature is less than $30,000$ K and whose
overdensity is greater than 1000.  We know that this approximately
picks out the neutral
HI gas owing to comparisons we made with calculations that used detailed
but computationally expensive self-shielding corrections.  The hydrogen fraction
by mass in these simulations is 0.76, so the masses should be multiplied by
this factor to get an upper limit to the HI masses.  This is only an upper
limit, since some of the hydrogen could be molecular.  At first the mass
function rises steadily towards lower redshift in all the models as more
gas condenses into galaxies and small galaxies merge into larger systems.
But then in the higher amplitude models, such as OCDM, CCDM, and SCDM, the
mass function ceases to rise as the gas is converted into stars.
In all cases the flattening of the mass functions at
$M_{gas} \la 10^{9.5} M_\odot$ is a result of the simulations'
finite mass resolution.

\begin{figure}
\plotone{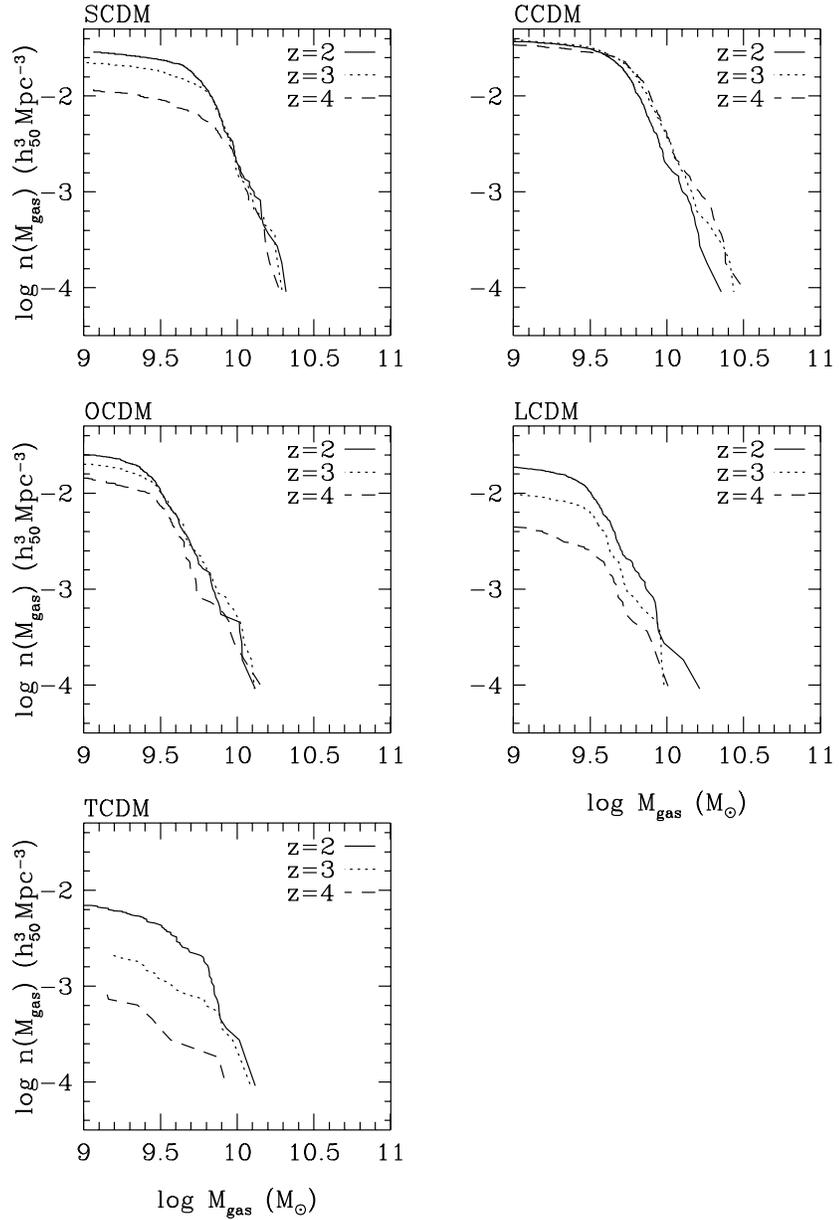}
\caption{Cumulative gas mass function in the different models at the
indicated redshifts: $n(M_{gas})$ is the comoving number density of galaxies
whose gas mass exceeds $M_{gas}$.} \label{fig-3}
\end{figure}

In Ingram et al. (1996) we ``observed'' simulations such as these with the
Square Kilometer Array Interferometer (SKAI) to make artificial
HI emission maps.  We determined that the SKAI should detect about 
$10^{10} \msun$/beam in a 100 hour observation at the $5\sigma$ level.
This can be compared to the galaxy gas masses in
Figure 3.  At these redshifts the beam should contain each galaxy.
In several of the models, galaxies should be detectable by SKAI.
By observing the gas mass function it may be possible to discriminate between
different cosmological models.

Besides measuring the gas mass function of galaxies it should also be possible
to measure the galaxy-galaxy correlation function.  These are plotted for
each model in Figure 4 at $z = 2$, 3, and 4. The solid lines are the
correlation functions for the dark matter.  As expected, the correlation
functions of the dark matter rise towards lower redshift as structure
grows in the universe.  However, the correlation amplitude of the galaxies
remains almost constant and in some cases even becomes smaller.  This
occurs because the galaxies are a highly biased population (Bagla 1997).
The bias of the galaxies, which is proportional to the square root of the
ratio of the correlation amplitude of the galaxies compared to that of the 
dark matter, ranges from 1.5 to 4 and is even higher when only the most massive
galaxies are considered.

We plan to continue using the simulations to make detailed artificial
HI maps, like those in Ingram et al. (1996), both to make testable
predictions of the models and to help guide the construction of future
large radio telescopes.  We also plan to make artificial CO maps for
comparisons with large mm telescopes such as the LMT.

\begin{figure}
\plotone{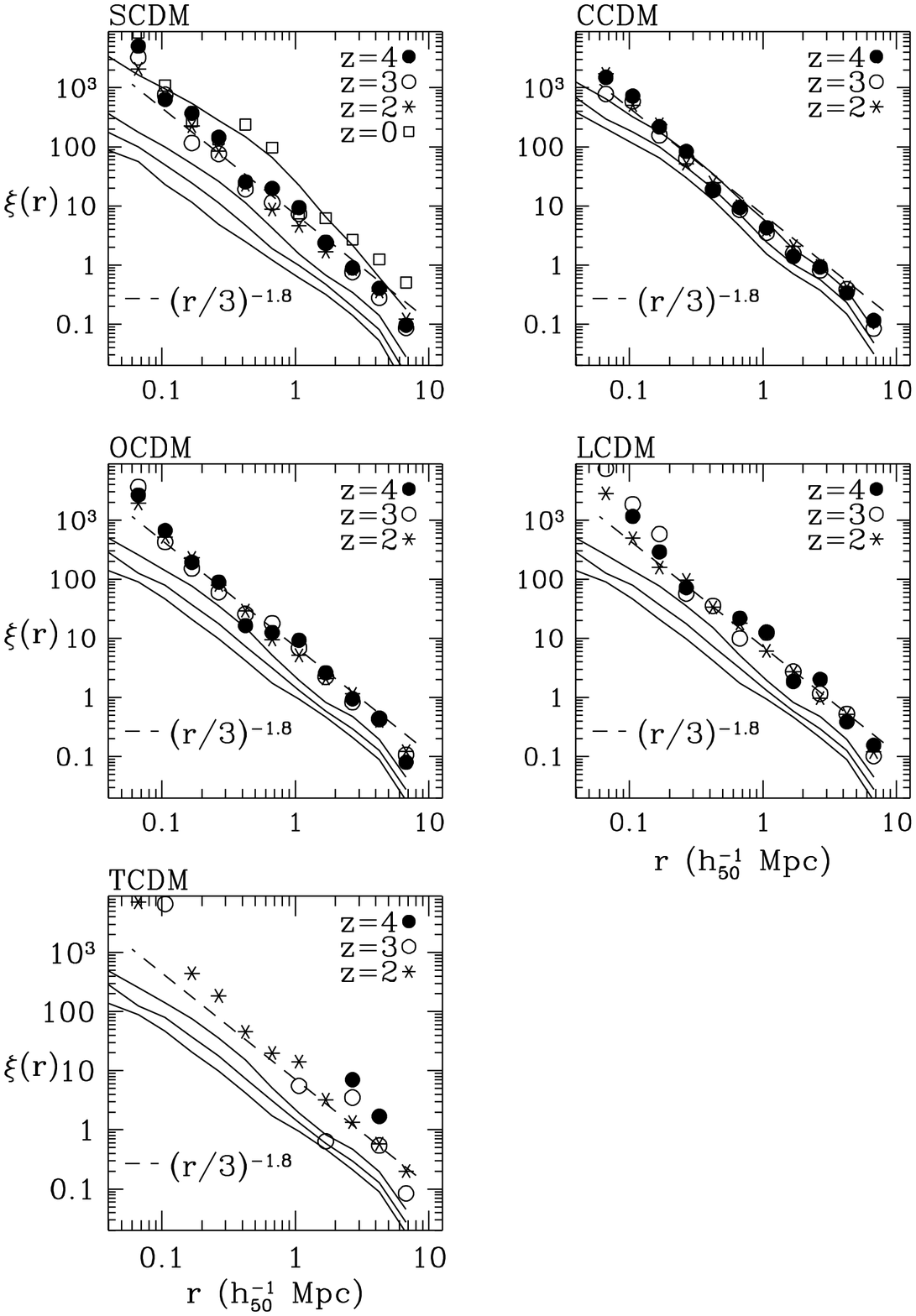}
\caption{Symbols show the galaxy-galaxy correlation functions at the indicated redshifts.  The solid lines are the correlation functions for the dark matter.
The dashed line is a power law with a slope of -1.8 and a correlation length of
$3\;h_{50}^{-1}\;{\rm Mpc}. $} \label{fig-4}
\end{figure}

\acknowledgments

We would like to thank Eric Linder for useful discussions.
This work was supported by NASA grants NAG5-3525, NAG5-3922, and NAG5-4064.

\end{document}